\pdfoutput=1

\documentclass[conference]{IEEEtran}
\IEEEoverridecommandlockouts
\usepackage{enumitem}
\usepackage{todonotes}
\usepackage{booktabs}
\usepackage{epstopdf}
\usepackage{graphicx}
\usepackage{listings}
\usepackage{multirow}
\usepackage{array}
\usepackage{siunitx}
\usepackage{subfigure}
\usepackage{mathtools}
\usepackage{url}
\usepackage[normalem]{ulem}

\newcommand{\upperRomannumeral}[1]{\uppercase\expandafter{\romannumeral#1}}

\ifCLASSOPTIONcompsoc
  \usepackage[nocompress]{cite}
\else
  \usepackage{cite}
\fi
\ifCLASSINFOpdf
\else
\fi
\usepackage{amsmath}

\DeclareMathOperator*{\argmin}{arg\,min}

\begin{document}
%
\title{One-Hop Out-of-Band Control Planes for Low-Power Multi-Hop Wireless Networks
\thanks{This work was supported in part by an NTU CoE Seed Grant, an NTU Start-up Grant, MOE Tier 2 under Grant MOE2014-T2-2-015 ARC4/15 and NRF2015-NRF-ISF001-2277.}}




\author{\IEEEauthorblockN{Chaojie Gu\IEEEauthorrefmark{1} $\qquad$
Rui Tan\IEEEauthorrefmark{1} $\qquad$ Xin Lou\IEEEauthorrefmark{2} $\qquad$
Dusit Niyato\IEEEauthorrefmark{1}}
\IEEEauthorblockA{\IEEEauthorrefmark{1}School of Computer Science and Engineering, Nanyang Technological University, Singapore}
\IEEEauthorblockA{\IEEEauthorrefmark{2}Advanced Digital Sciences Center, Illinois at Singapore}
}


\maketitle
\sloppy
\begin{abstract}

Separation of control and data planes (SCDP) is a desirable paradigm for low-power multi-hop wireless networks requiring high network performance and manageability. Existing SCDP networks generally adopt an {\em in-band} control plane scheme in that the control-plane messages are delivered by their data-plane networks. The physical coupling of the two planes may lead to undesirable consequences. To advance the network architecture design, we propose to leverage on the long-range communication capability of the increasingly available low-power wide-area network (LPWAN) radios to form one-hop out-of-band control planes. We choose LoRaWAN, an open, inexpensive, and ISM band based LPWAN radio to prototype our out-of-band control plane called LoRaCP. Several characteristics of LoRaWAN such as downlink-uplink asymmetry and primitive ALOHA media access control (MAC) present challenges to achieving reliability and efficiency. To address these challenges, we design a TDMA-based multi-channel MAC featuring an urgent channel and negative acknowledgment. On a testbed of 16 nodes, we demonstrate applying LoRaCP to physically separate the control-plane network of the Collection Tree Protocol (CTP) from its ZigBee-based data-plane network. Extensive experiments show that LoRaCP increases CTP's packet delivery ratio from 65\% to 80\% in the presence of external interference, while consuming a per-node average radio power of $2.97\,\text{mW}$ only.

\end{abstract}


%
\IEEEpeerreviewmaketitle

\section{Introduction}
\label{sec:intro}
Billions of smart objects will be deployed, forming {\em things networks} that are interconnected by Internet of Things (IoT). Many of these networks will follow the multi-hop wireless paradigm. For instance, wireless meshes are increasingly adopted to interconnect surveillance cameras \cite{city-mesh} and vehicles \cite{veniam}. Wireless sensors have been widely deployed for sensing and control of building environment and energy use. Bluetooth low energy (BLE) will support mesh networking soon \cite{ble-mesh}. Wireless connectivity is also critical to the vision of Industry 4.0. Utility and manufacturing systems are increasingly adopting wireless metering and monitoring \cite{wirelesshart}.

The main advantage of \underline{l}ow-power \underline{m}ulti-hop \underline{w}ireless \underline{n}etworks (LMWNs) is that, during the deployment phase, a network can easily scale up to cover a large geographic area. A primary design principle for LMWNs is the use of distributed protocols (e.g., routing \cite{gnawali2009collection}), where each node independently performs various networking functions (e.g., data forwarding) based on local information. Thus, the {\em control plane} (i.e., determination of how to handle packets) and the {\em data plane} (i.e., carrying out control-plane decisions) of these distributed protocols are jointly implemented at each network node. However, as a well understood notion, a distributed scheme without the global view often yields suboptimal performance. Moreover, although the distributed scheme may work satisfactorily most of the time thanks to a decade of research, it is often complex, inelastic to change, and difficult to manage once the network is deployed.

To improve the network performance and manageability, some LMWNs, especially those deployed for mission-critical tasks, have adopted centralized network controls. For instance, WirelessHART, an LMWN standard that has been adopted in over 8,000 manufacturing systems \cite{wirelesshart}, prescribes centralized routing control based on a global view of the network to better achieve certain performance objectives (e.g., firm/soft real-time packet delivery). Similarly, ISA100.11a, another industry-oriented LMWN standard, also adopts centralized routing control and network management. For the routing in these LMWNs, the control plane is separated from the data plane, in that the routing control is implemented at a centralized node whereas other network nodes follow the routing schedule to forward data packets. However, all these LMWNs adopt {\em in-band} control planes, i.e., the control-plane messages such as network status reports and routing schedules are delivered by the data-plane networks.

The physical coupling between the control and data planes in the in-band scheme may lead to undesirable consequences. The wireless data-plane network is susceptible to external interference. Deteriorated data-plane links may lead to delayed deliveries or even losses of the control-plane messages, making the network less responsive to data-plane link quality variations. Moreover, when the data plane loses key routing nodes (e.g., due to node hardware/software fault and depletion of battery) or the control plane makes wrong control decisions (e.g., due to design defects or erroneous human operations), the data-plane network may fall apart to disconnected partitions. As a result, restorative network control commands in the control plane may not be able to reach the destination nodes. Recent research has studied protecting the control plane from data-plane faults \cite{huang2016near}. However, the solution has limited protection capability against a single link failure only \cite{huang2016near}.

In light of the in-band scheme's pitfalls, we study an {\em out-of-band} scheme, where the control plane uses a dedicated network different from the data-plane network. The increasing availability of multiple network interfaces on IoT hardware platforms favors the implementation and adoption of the out-of-band scheme. The latest IoT platforms are generally equipped with multiple heterogeneous network interfaces: Raspberry Pi 3 supports Ethernet, Wi-Fi, and BLE; Firestorm \cite{andersen2016system} supports BLE and ZigBee; Arduino has various add-on boards to support different radios. To design the out-of-band control-plane network for LMWN, most high-speed radios (e.g., Wi-Fi and LTE) are unsuitable due to their high power consumption. ZigBee and the coming BLE mesh are also ill-suited, since otherwise the control-plane network will be yet another LMWN that suffers the same manageability and fragility issues as the data-plane network. Instead, we propose to use the emerging low-power wide-area network (LPWAN) technologies (e.g., LoRaWAN, SigFox, Weightless-P, and NB-IoT) for the out-of-band control plane. Owing to the kilometers communication range of LPWAN links, the LPWAN-based control plane can be a one-hop star network, greatly simplifying its deployment and management.

As the first study to our best knowledge on the feasibility of LMWN out-of-band control plane, we choose LoRaWAN to prototype our system and gain insights. This choice is due to its use of license-free ISM band, open data link standard, low cost (US\$15 per unit \cite{sx1276}), and good scalability to support many IoT objects.
In contrast, other LPWAN technologies are proprietary (SigFox and NB-IoT) or not widely available (Weightless-P). While the low-power long-range communication capability is the key advantage of LoRaWAN, we need to manage the following two limiting characteristics of LoRaWAN. First, a LoRaWAN downlink frame from the controller to a network node must be in response to a precedent uplink frame. Thus, the transmissions of network control commands initiated by the controllers may be postponed to the network node's status reporting. Second, LoRaWAN supports uplink concurrency but no downlink concurrency. This downlink-uplink asymmetry impedes acknowledging each uplink frame, whereas the control plane generally desires reliable message delivery. In addition, a reliable media access control (MAC) approach is needed to replace LoRaWAN's ALOHA MAC that may perform unsatisfactorily in traffic surges.

To address these issues, this paper presents the design and implementation of a prototype system called {\em LoRaCP} (\underline{lo}ng-\underline{ra}nge \underline{c}ontrol \underline{p}lane). Based on our extensive measurements on LoRaWAN's energy and latency profiles, we design {\em LoRaCP-MAC}, a TDMA-based multi-channel MAC protocol featuring uplink heartbeats, negative acknowledgment (NAK), and an ALOHA-based urgent channel, to manage the transmissions of the control-plane messages. The uplink heartbeats open downlink windows for controller-initiated network commands and maintain network nodes' clock synchronization for TDMA. With NAK, the controller needs not acknowledge every uplink frame. The urgent channel complements the TDMA channels to mitigate the rigidness of TDMA. On a testbed of 16 nodes, we demonstrate applying LoRaCP to physically separate the control plane of the Collection Tree Protocol (CTP) \cite{gnawali2009collection} from its ZigBee-based data-plane network. Extensive experiments show that LoRaCP increases CTP's packet delivery ratio from 65\% to 80\% in the presence of external interference, while consuming a per-node average radio power of $2.97\,\text{mW}$ only, much lower than the active power of many recent LMWN platforms' microcontrollers (e.g., $28.38\,\text{mW}$ on Firestorm \cite{andersen2016system}).

The rest of the paper is organized as follows. \S\ref{sec:related} reviews related work. \S\ref{sec:motivating} presents examples to motivate the out-of-band scheme. \S\ref{sec:profiling} profiles LoRaWAN performance. \S\ref{sec:design} and \S\ref{sec:eval} design and evaluate LoRaCP, respectively. \S\ref{sec:conclude} concludes.
\section{Related Work}
\label{sec:related}
Existing studies that exploit multiple network interfaces can be broadly divided into two classes of {\em bandwidth aggregation} and {\em separation of control and data planes} (SCDP).

Bandwidth aggregation uses multiple network interfaces to transmit/receive data simultaneously to increase throughput. Habak et al. \cite{habak2015bandwidth} surveyed early bandwidth aggregation literature. Recent development is reviewed briefly here. We divide them into two categories. The first category exploits homogeneous radios. FatVAP \cite{kandula2008fatvap} enables a 802.11 wireless card to connect to multiple access points. FastForward \cite{ekbatanifard2013fastforward} uses two 802.15.4 radios operating on different channels, with one receiving and the other forwarding data simultaneously. The second category exploits heterogeneous radios. MultiNets \cite{nirjon2014multinets} deals with the switching between multiple network interfaces on mobile devices. In \cite{mu2017iwqos}, Mu et al. optimize the selection of radios and their transmission powers. Recent studies \cite{lim2015design,nikravesh2016depth} characterize the performance and energy consumption of Multipath TCP through multiple radios of a mobile device. Different from bandwidth aggregation that combines multiple network interfaces in the data plane to increase throughput, SCDP aims to improve network optimality and manageability.

Software-defined networking (SDN), with SCDP as its core concept, is a growing momentum in data-intensive networks. SCDP can be naturally applied in WLANs and cellular networks, as their topologically centralized access points and base stations can run the control-plane logics for better resource allocation and mobile node handover \cite{jagadeesan2015software}. However, there is limited research on SCDP in multi-hop wireless networks. An OpenFlow-enabled Wi-Fi mesh was built in \cite{dely2011openflow}, where each Wi-Fi card is split into two virtual interfaces with different SSIDs and the two planes are two multi-hop networks in their respective SSIDs. To the best of our knowledge, WASP \cite{kaplan2014wasp} is the only system that implements out-of-band control plane for multi-hop wireless networks. WASP uses Wi-Fi Direct and LTE of smartphones to form the data and control planes, respectively. Different from WASP, we focus on low-power networks with a limited energy budget.

\section{Motivation}
\label{sec:motivating}

This section discusses the motivation of the out-of-band scheme. \S\ref{subsec:distributed-vs-centralized} presents a simulation study to show the network performance gain by centralized network control. In \S\ref{subsec:pitfalls}, we discuss the challenges faced by the in-band scheme.

\subsection{Distributed vs. Centralized Network Control}
\label{subsec:distributed-vs-centralized}

In this section, we compare through simulations the network performance achieved by the Collection Tree Protocol (CTP) \cite{gnawali2009collection} and its centralized variant that we call CTP-SCDP. In \S\ref{sec:eval}, we will use LoRaCP to implement CTP-SCDP and evaluate it on a testbed. In this paper, we use CTP as our case study network protocol, because it has an open implementation and is a standard component of the industry-class TinyOS Production operating system \cite{tinyos-prod}. The results based on CTP will help understand the performance improvement by SCDP and showcase the use of LoRaCP to physically separate the control and data planes. We believe the understanding and LoRaCP are also applicable to many other LMWN protocols.

CTP aims to maintain a minimum-cost routing tree in the presence of dynamic link quality characterized by the expected transmission count (ETX). The cost of a route to the tree root is the sum of the ETXs of the links on the route. A node $i$ estimates the route cost using the residual ETX (RETX), which is given by $\mathrm{RETX}_i = \mathrm{ETX}_{i,p} + \mathrm{RETX}_{p}$, where $\mathrm{ETX}_{i,p}$ is the ETX of the link between node $i$ and its parent node $p$, and $\mathrm{RETX}_p$ is node $p$'s RETX. CTP works in a distributed manner, in that each node $i$ selects its parent $p$ from the set of its neighbor nodes $\mathcal{N}$ based on local information only. Specifically, $p = \argmin_{j \in \mathcal{N}} \mathrm{ETX}_{i,j} + \mathrm{RETX}_{j}$, where $\mathrm{ETX}_{i,j}$ is estimated based on the transmissions of beacons and data frames; $\mathrm{RETX}_{j}$ is broadcast in node $j$'s beacons.

\begin{figure}
  \centering
  \includegraphics{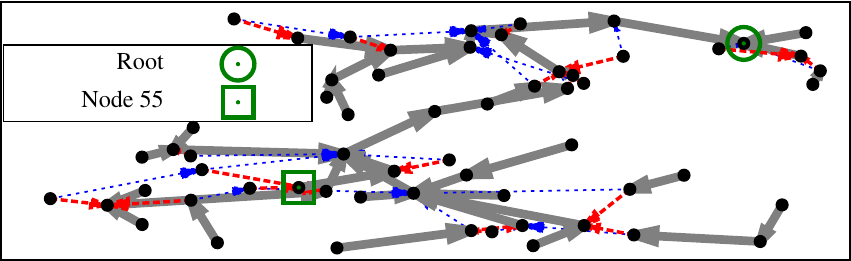}
  \vspace{-0.5em}
  \caption{Routing trees by CTP and CTP-SCDP. The solid thick gray links are shared by the CTP and CTP-SCDP trees; the dashed thick red links are on the CTP tree only; the dashed thin blue links are on the CTP-SCDP tree only.}
  \label{fig:trees}
\end{figure}

In CTP, the information about the quality of a link propagates to the whole network during the beaconing process. However, this propagation takes time. Thus, when link quality changes over time, the RETX of any node $i$ cannot capture the latest ETXs of the links on its route to the root. In particular, the closer the links on the route are to the root, node $i$'s knowledge about the links (which is encompassed in $\mathrm{RETX}_i$) is more out-of-date. As a result, CTP may not construct the minimum-cost tree in the presence of time-varying link quality. Differently, in CTP-SCDP, the latest ETXs are updated to the network controller and the optimal routing is sent to the nodes, both directly through the control-plane network.

We compare CTP and CTP-SCDP using the TinyOS simulator TOSSIM. We place 60 nodes randomly in a $200\,\text{m} \times 200\,\text{m}$ region as illustrated in Fig.~\ref{fig:trees}. Link gains are generated according to the Euclidean distances between nodes using a tool in TOSSIM. Radios' hardware noise floor is set to be $-90\,\text{dBm}$ (a mild noise level). To simulate CTP-SCDP, we add a node as the network controller, which has sufficiently large link gains with any other nodes, such that the control-plane network is a one-hop star network. In CTP-SCDP, node $i$ sends the latest $\mathrm{ETX}_{i,j}$ to the network controller. Upon receiving an ETX update, the controller updates a directed graph with the ETXes as the edge costs and recomputes the minimum-cost routing tree using the Dijkstra's algorithm. Then, the controller sends the new parent information to the nodes.

\begin{figure}
  \subfigure[Node 55]
  {
    \includegraphics{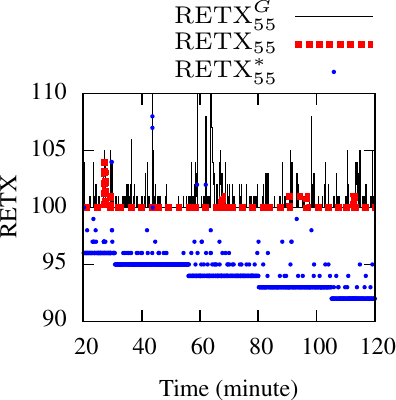}
    \label{fig:ctp-performance-node-55}
  }
  \subfigure[All nodes]
  {
    \includegraphics{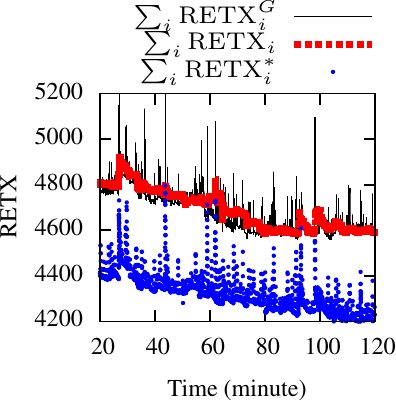}
    \label{fig:ctp-performance-all-nodes}
  }
  \vspace{-0.5em}
  \caption{CTP cannot build a minimum-cost tree.}
  \label{fig:ctp-performance}
\end{figure}

\begin{figure}
  \subfigure[Node 55]
  {
    \includegraphics{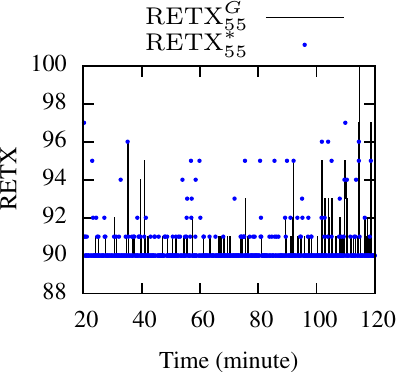}
    \label{fig:ctp-scdp-performance-node-55}
  }
  \subfigure[All nodes]
  {
    \includegraphics{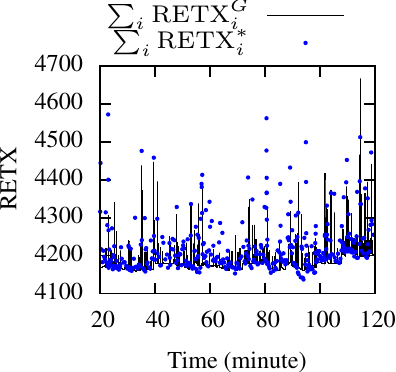}
    \label{fig:ctp-scdp-performance-all-nodes}
  }
  \vspace{-0.5em}
  \caption{CTP-SCDP can build a minimum-cost tree.}
  \label{fig:ctp-scdp-performance}
  \vspace{-0.5em}
\end{figure}

We conduct two sets of simulations to show the benefits of SCDP. The first set shows the suboptimal performance of CTP. Specifically, we concurrently run CTP and CTP-SCDP, but the controller in CTP-SCDP does not send routing control commands to the nodes. Thus, the routing is managed by CTP only. We consider the following evaluation metrics:
\begin{enumerate}
\item RETX of node $i$ estimated by CTP (denoted by $\mathrm{RETX}_i$) and the sum of all RETXes (denoted by $\sum_{i} \mathrm{RETX}_i$);
\item The {\em ground-truth} RETX of the route determined by CTP for node $i$ (denoted by $\mathrm{RETX}_i^G$), which can be measured as the sum of the latest ETXes of the links on the route obtained by the controller in CTP-SCDP, as well as the sum of all ground-truth RETXes (i.e., $\sum_i\mathrm{RETX}_i^G$);
\item The minimum RTEX of node $i$ computed by CTP-SCDP (denoted by $\mathrm{RETX}_i^*$) and the sum $\sum_i\mathrm{RETX}_i^*$.
\end{enumerate}
We simulate a time duration of two hours, during which each node generates a data packet every eight seconds. Fig.~\ref{fig:trees} shows the routing trees computed by CTP and CTP-SCDP at the end of the simulation. They are different. Fig.~\ref{fig:ctp-performance} shows the evaluation metrics for Node~55 and all the nodes over the two hours. We can see that, compared with the ground truth (i.e., the solid black curves), CTP's knowledge about the chosen routes (i.e., the dashed red curves) cannot capture many transient changes in the ground truth, because of the information propagation latency in the distributed network control. Compared with the global optimal (i.e., the blue dots), the routes chosen by CTP have higher costs.

In the second set of experiments, we run CTP-SCDP only. Fig.~\ref{fig:ctp-scdp-performance} shows the results. From the figure, we can see that the routes chosen by CTP-SCDP generally achieve the minimum costs. The above two sets of simulations show that centralized network control improves the network performance in dynamic network conditions. Thus, the centralized control enabled by SCDP is desirable for performance-critical networks such as those deployed for industrial applications \cite{wirelesshart}.

\subsection{Challenges Faced by In-Band Control Planes}
\label{subsec:pitfalls}

Our simulations in \S\ref{subsec:distributed-vs-centralized} demonstrate the underperformance of distributed network control. As discussed in \S\ref{sec:intro}, some mission-critical LMWNs have adopted centralized network control to improve network performance and manageability. However, they follow the in-band control plane scheme due to the lack of multiple radios in the past decade or the concern of increased infrastructure cost in deploying additional radios. However, the physical coupling of the control and data planes generates various challenges. For instance, given the fragile nature of wireless, how to protect the in-band control plane against data-plane link failures is a challenging problem. Recent research has investigated this issue. However, existing solutions provide limited protection capability. For instance, the solution proposed in \cite{huang2016near}, though sophisticated, can handle a single link failure only. The in-band control plane protection under a general setting is still an open issue.

Given the complications of the in-band scheme and the resulted, unsolved challenges, in this paper, we study the alternative out-of-band scheme that is increasingly feasible in terms of hardware support, due to the prevalence of multiple radios on IoT platforms. In particular, LPWAN radios are becoming readily available and cheap (US\$15 per unit \cite{sx1276}). Thus, we inquire in this paper basic system research questions including the feasibility and design of LPWAN-based control plane for LMWNs, as well as its performance under various settings. In the following sections, the design and evaluation of LoRaCP provide a baseline in answering these questions.
\section{LoRaWAN Performance Profiling}
\label{sec:profiling}
This section profiles LoRaWAN's energy and latency, which are important to the design of LoRaWAN-based control planes.

\subsection{LoRaWAN and Its Characteristics}
\label{subsec:intro-lorawan}

\subsubsection{Introduction of LoRaWAN}

LoRaWAN (long-range wide area network) is an open data link layer specification based on LoRa, a PHY layer technique that uses a Chirp Spread Spectrum modulation and operates in sub-GHz ISM bands (e.g., EU$868\,\text{MHz}$ and US$915\,\text{MHz}$). LoRa admits configuring the ratio between the symbol rate and chip rate by specifying an integer {\em spreading factor} (SF) within $[6, 12]$. Specifically, each symbol is modulated by $2^\text{SF}$ chips. A higher SF increases the signal-to-noise ratio and the communication range, but decreases the symbol rate. In this paper, we use six SF settings, i.e., from SF7 to SF12. (SF6 is a special setting that is often not used.) The communications using different SFs are orthogonal and thus can be concurrent. Thus, in this paper, the terms {\em SF} and {\em channel} are used interchangeably.

A LoRaWAN network is formed by one or more {\em gateways} and many {\em end devices}. The gateway, often Internet-connected, can simultaneously handle the communications with multiple nodes in different channels. LoRaWAN defines three classes (A, B, and C) of end devices. A Class-A device's uplink transmission is followed by two downlink windows (RX1 and RX2). Downlink communications to the node at any other time will have to wait until the next uplink from the node. As Class-A is the most power efficient and supported by any end device, we choose to design LoRaCP based on Class-A. The Class-B and C have not been widely supported. Designing LoRaCP based on them will decrease LoRaCP's universality.

\subsubsection{Characteristics of LoRaWAN}
\label{subsubsec:characteristics}

The low-power long-range communication capability is the main advantage of LoRaWAN that makes it promising for control planes of LMWNs. However, we need to keep in mind the following two limiting characteristics of LoRaWAN in the design of LoRaCP.

\noindent
{\bf Downlink-uplink asymmetry:} LoRaWAN is mainly designed and optimized for uplinks from end devices to gateway. For instance, the LoRaWAN concentrator can receive frames from multiple channels simultaneously, whereas it can send a single downlink frame only at a time. Moreover, the Class-A specification requires that any downlink transmission must be unicast, in response to a precedent uplink transmission.

\noindent
{\bf Lossy links:} From existing tests \cite{marcelis2017dare}, with SF12, the frame reception rate is about 80\% at a distances of $2.5\,\text{km}$. To build a reliable control-plane network, the frame losses need to be dealt with properly. Acknowledging each uplink frame is wasteful given the scarce downlink time as discussed earlier.

By default, LoRaWAN uses ALOHA that may perform unsatisfactorily in surges of control plane messages. Thus, we need to design a new MAC to enable efficient LoRaCP. As LoRa does not prescribe carrier sense capability, CSMA is not viable. Time-division multiple access (TDMA) is often adopted for reliability that control planes desire. However, as shown in this paper, the implementation of TDMA on LoRaWAN is non-trivial. Moreover, a strict TDMA may result in undesirable delays in transmitting urgent messages.	

In the design of LoRaCP (cf.~\S\ref{sec:design}), the downlink-uplink asymmetry and lossy links will be managed by the NAK mechanism. Moreover, we will design a TDMA-based multi-channel MAC with an urgent channel to replace the ALOHA. Although we face the above limiting characteristics of LoRaWAN, this work presents software solutions that can be implemented readily on various LMWN platforms that integrate LoRa radios. Our software-space design is much more cost effective and practical than clean slate LPWAN hardware designs for LMWN control planes.
	
\subsection{LoRaWAN Performance Profiling}
\label{subsec:lorawan-profiling}

\subsubsection{LoRaCP hardware prototypes}
\label{subsubsec:hardware-prototypes}

We conduct performance profiling based on the following prototype hardware platforms. Each end device integrates a Cooking Hacks LoRaWAN shield \cite{lorawan-shield} and a Raspberry Pi (RPi) 3 Model B single-board computer. The shield has a Microchip RN2483 LoRaWAN chip, an $868\,\text{MHz}$ antenna, and interfacing circuits. The shield can be controlled by the RPi using a C++ library from Cooking Hacks. The gateway integrates an RPi and an IMST iC880A LoRaWAN concentrator board \cite{ic880a}. The iC880A board can receive frames over all LoRa channels simultaneously.

\begin{figure}
  \centering
  \subfigure[LoRaCP node]
  {
    \includegraphics[width=.15\textwidth]{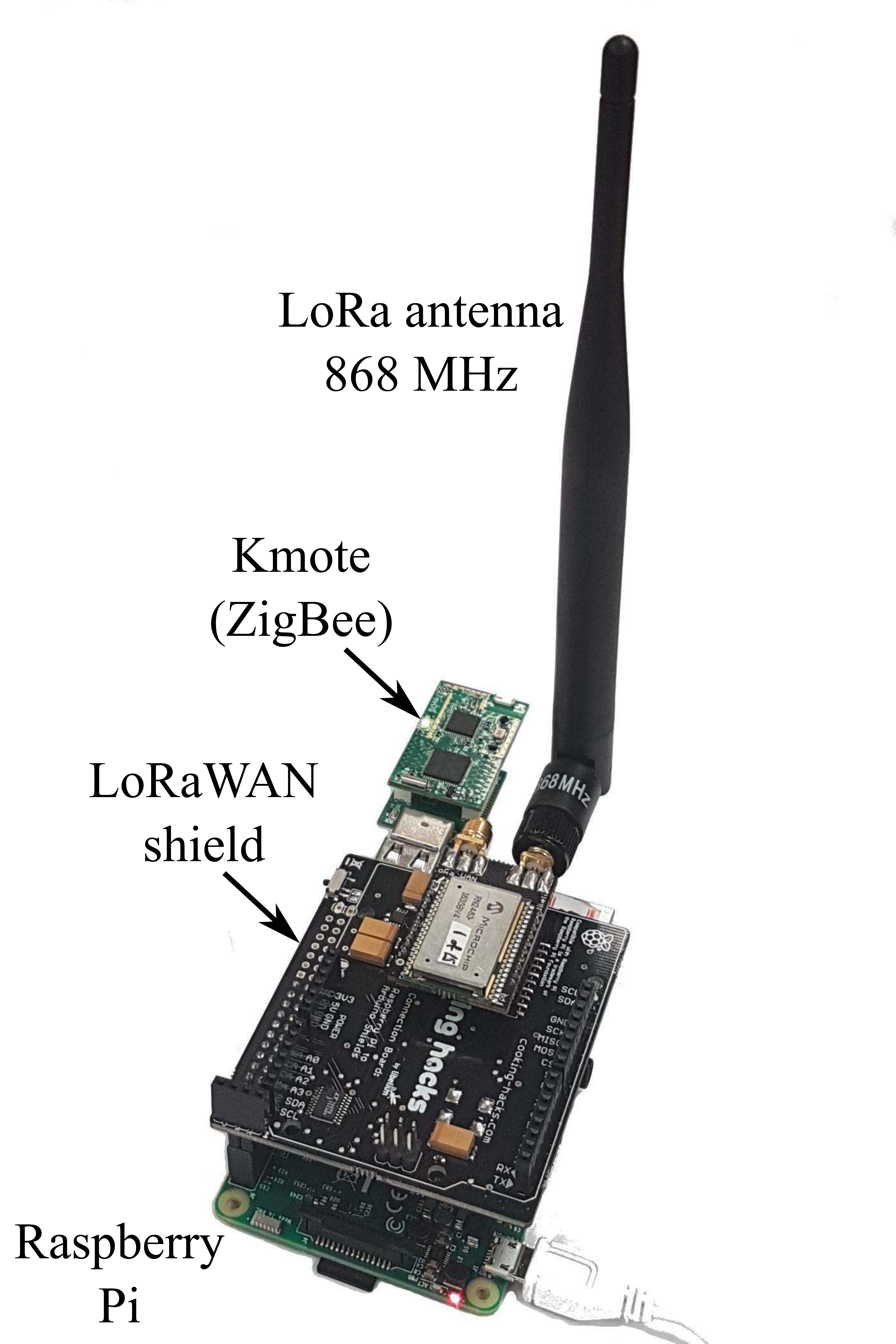}
  }
  \hspace{1em}
  \subfigure[LoRaCP controller]
  {
    \includegraphics[width=.15\textwidth]{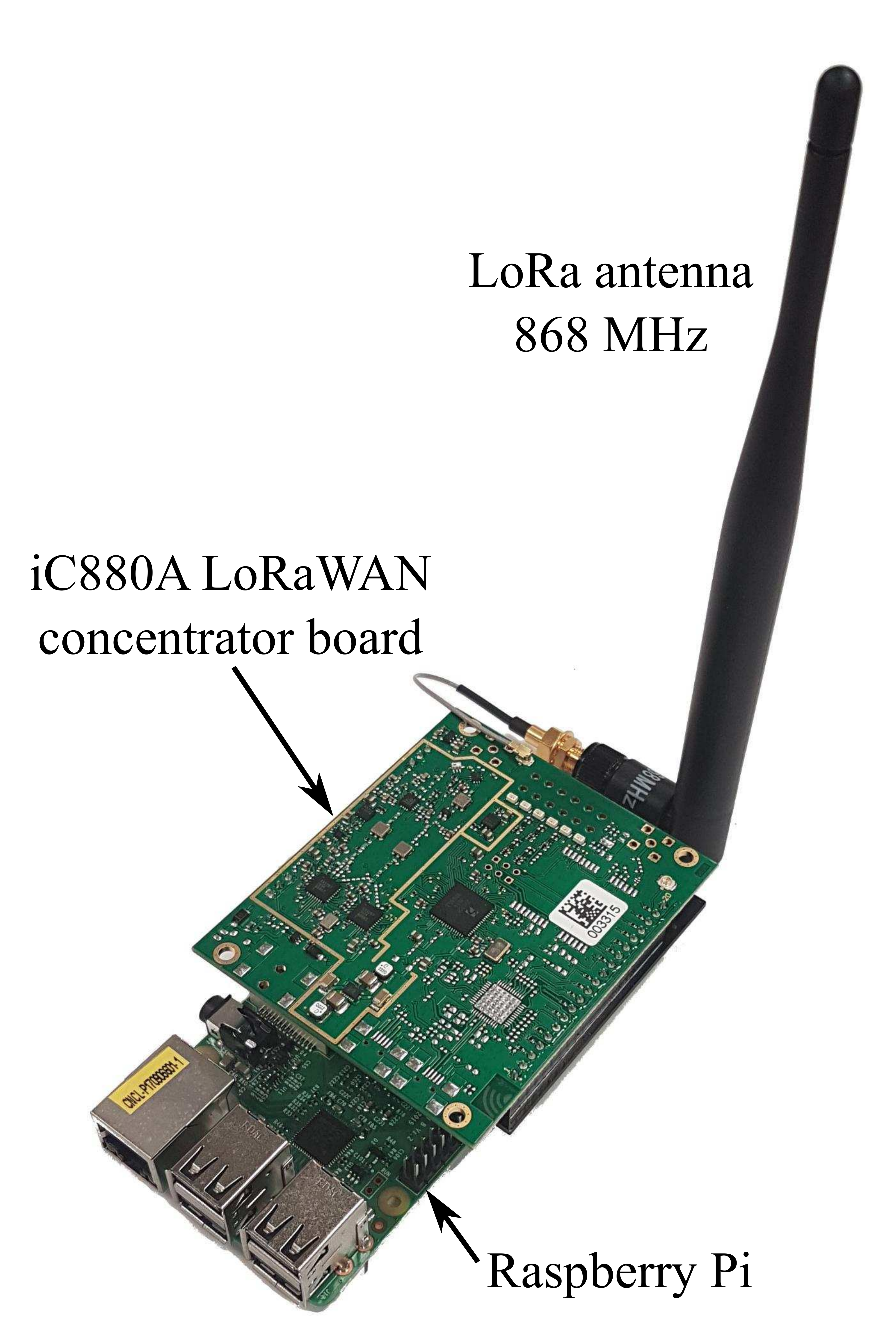}
  }
  \caption{LoRaCP hardware prototypes. (The Raspberry Pi is for fast prototyping only; it will not be needed if LoRa is built into the LMWN platform.)}
  \label{fig:prototype}
\end{figure}

A ZigBee-based Kmote is plugged into a USB port of the RPi of each end device, forming a {\em LoRaCP node}. The nodes use their ZigBee radios to form the data-plane network. From now on, the gateway is referred to as {\em LoRaCP controller}. The controller unnecessarily has a ZigBee radio, since it may not be in the data-plane network. We use RPi to quickly prototype the integration of LoRaWAN and ZigBee. The results of this paper will suggest that integrating LoRaWAN into the design of LMWN platforms, especially those desiring high network performance and manageability, is valuable. In such designs, the RPi will not be needed. Fig.~\ref{fig:prototype} shows our prototypes.

\subsubsection{Energy profiling}

RN2483's datasheet says its run-time current supply is $38.9\,\text{mA}$ at $3.3\,\text{V}$. We use a Monsoon meter to measure the current supply of the whole LoRaWAN shield after properly jumping the power wires. Monsoon reads $39.5\,\text{mA}$ to $40.4\,\text{mA}$ under different SFs. This shows that the shield's encapsulating and interfacing circuits consume little power. The current supplies in the receiving and sleep modes are $14.2\,\text{mA}$ and $0.0016\,\text{mA}$, respectively.

A possible concern about LoRaWAN is its low data rate to power consumption ratio (DPR), compared with other low-power radios. For instance, with SF7 in the EU$868\,\text{MHz}$ band, the DPR is $11\,\text{kbps}/38.9\,\text{mA}=0.28\,\text{kbps/mA}$. In contrast, the DPR for ZigBee is $250\,\text{kbps}/19.5\,\text{mA}=12.82\,\text{kbps/mA}$. However, the severity of this concern should be discriminated regarding the aimed communication range. We illustrate this by an example of moving $x$ bits of data over a distance of $L$ meters by multiple hops. The radio energy used to move the $x$ bits over a hop is $(P_{Tx} + P_{Rx})\cdot \frac{x}{v}$, where $P_{Tx}$ and $P_{Rx}$ are the transmitting and receiving powers, respectively; $v$ is the link data rate in bps. Thus, the total energy used by the network's radios to move $x$ bits over $L$ meters is $(P_{Tx} + P_{Rx})\cdot \frac{x}{v} \cdot \frac{L}{d}$, where $d$ is the typical one-hop transmission range. Considering $L=1\,\text{km}$, we set $\frac{L}{d}$ to be 1 and 10 for LoRaWAN and ZigBee, respectively. Moreover, we set the data rate $v$ to be $11\,\text{kbps}$ and $250\,\text{kbps}$ for LoRaWAN and ZigBee, respectively. After applying respective power consumption measurements, LoRaWAN's total radio energy consumption is 2.94 times of ZigBee's. Although the above simplistic energy consumption estimation does not consider other factors like nodes' processor energy consumption and MAC, the result underlines our understanding. While LoRaWAN consumes more energy than ZigBee, it substantially simplifies the control-plane network design due to its one-hop nature. Moreover, the concern of LoRaWAN's higher energy consumption can be mitigated by the fact that the control plane's traffic volume is much lower than the data plane's. For instance, as measured in \S\ref{sec:eval}, the number of CTP-SCDP's control-plane frames is just about 5\% of its data-plane packets. Thus, we believe that, for the control-plane networks, the energy saving by using high-DPR but short-range radios is not worth sacrificing network simplicity.

\subsubsection{Latency profiling}
\label{subsec:lora-latency}

\begin{figure}
  \centering
  \begin{minipage}[t]{.23\textwidth}
    \includegraphics{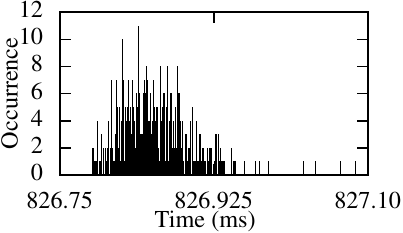}
    \caption{Radio awaking latency.}
    \label{fig:node_turn_on_time}
  \end{minipage}
  \hspace{1em}
  \begin{minipage}[t]{.22\textwidth}
    \includegraphics{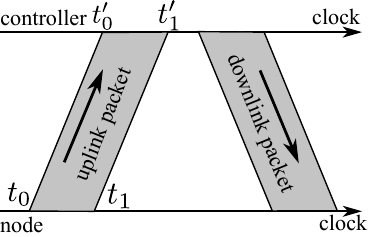}
    \caption{A communication session.}
    \label{fig:lora_ntp}
  \end{minipage}
  \vspace{-0.5em}
\end{figure}

Under TDMA, the LoRaWAN radio can sleep to save energy while waiting for the next time slot. The time delays in awaking the radio and transmitting a frame are critical to the radio's sleep scheduling and clock synchronization required by TDMA, respectively. We measure the latency in awaking the radio from the RPi using the shield's C++ API. Fig.~\ref{fig:node_turn_on_time} shows the distribution of the awaking latency over 500 tests. The mean and standard deviation are $826.9\,\text{ms}$ and $0.044\,\text{ms}$, respectively. The small standard deviation suggests that a LoRaCP node can awake the radio punctually for the next TDMA time slot.

\begin{figure}
	\centering
	\includegraphics[width=.4\textwidth]{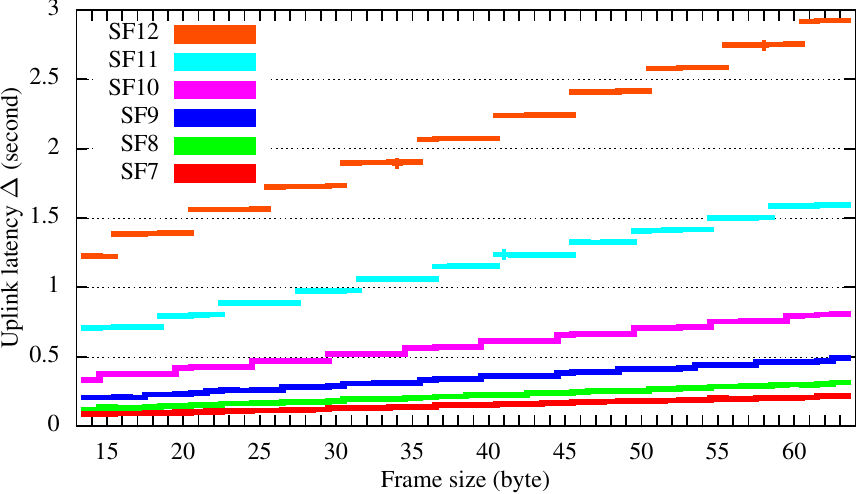}
        \vspace{-0.5em}
	\caption{Uplink latency under different SFs and frame sizes.}
	\label{fig:on_air_time}
\end{figure}

Then, we measure the latency in transmitting an uplink frame. Fig.~\ref{fig:lora_ntp} illustrates the uplink transmission's timing. The node starts and completes the transmission when its clock values are $t_0$ and $t_1$, respectively. The controller starts and completes the reception when its clock values are $t_0'$ and $t_1'$, respectively.
We can record $t_0$, $t_1$, and $t_1'$ in the LoRaWAN shield's and concentrator's C++ user programs running at their RPis. To measure the uplink latency, we synchronize the clocks of the node's and controller's RPis using the Network Time Protocol (NTP) over an Ethernet that gives sub-ms synchronization accuracy. We define the uplink latency as $\Delta = t_1' - t_0$.\footnote{We do not use $t_1$ because it contains a non-negligible uncertain delay from the actual completion of the transmission to the LoRaWAN shield's C++ library's periodic pull of the event from the shield's hardware interface.} Thus, the latency is determined by the data rate, which further depends on SF, and the frame size. Fig.~\ref{fig:on_air_time} shows the box plots of the measured uplink latency under different SFs and frame sizes. As the latency has little variations under each setting, the boxes and whiskers of the plots are not visible. We can see that the latency increases with both frame size and SF, which are consistent with our understanding. Interestingly, for a certain SF, the latency exhibits step changes when the frame size increases. This is because each LoRa frame is a certain number of bits aligned for easy hardware handling. The above measurement results lay a foundation for developing LoRaWAN clock synchronization in \S\ref{subsubsec:clock-sync}.

\section{Design and Implementation of LoRaCP}
\label{sec:design}
\subsection{System Overview and LoRaCP-MAC}
\begin{figure}
  \centering
  \includegraphics{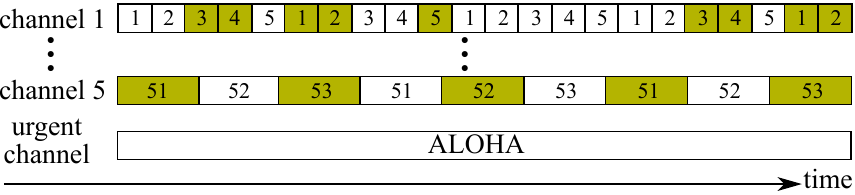}
  \caption{Illustration of LoRaCP-MAC. The number in a slot represents the ID of the node assigned to the slot. Shaded slots mean heartbeat slots.}
  \label{fig:LoRaCP-mac}
\end{figure}

The goal of LoRaCP is to use LoRaWAN's uplinks and downlinks to transmit {\em network reports} from the nodes to the controller and {\em network commands} from the controller to the nodes, respectively. The network commands have two categories: a {\em reactive network command} to a node is in response to a precedent network report from the node, whereas an {\em active network command} is initiated by the controller. All the control-plane transmissions are managed by LoRaCP's MAC protocol as illustrated in Fig.~\ref{fig:LoRaCP-mac}, which we call {\em LoRaCP-MAC}. As discussed in \S\ref{subsec:intro-lorawan}, LoRaWAN has six concurrent uplink channels. Five of them use TDMA, while the remaining one (called {\em urgent channel}) uses ALOHA to transmit urgent frames. The five concurrent TDMA channels increase the throughput for the network reports. The urgent channel mitigates the rigidness of TDMA and allows the control-plane application developers to deal with urgent situations such as sudden strong interference or even malicious jamming to the data-plane network. As the TDMA channels have different data rates, their time slot lengths can be different to achieve the same maximum frame size. The time slots of a TDMA channel are allocated in a round-robin fashion to the LoRaCP nodes that use the channel. The LoRaCP nodes can be assigned to the TDMA channels to balance their time delays in waiting for the next time slot, while considering the channels' communication ranges and the nodes' distances to the controller.

We now present two features of LoRaCP-MAC that address LoRaWAN's downlink-uplink asymmetry and lossy links.

\noindent
{\bf Heartbeat time slots:} When a node has no uplink data to transmit, it can skip its next time slot. However, because any downlink frame must be in response to a precedent uplink frame, LoRaCP-MAC designates periodic heartbeat time slots for each node. For instance, in Fig.~\ref{fig:LoRaCP-mac}, the shaded blocks represent heartbeat slots. In Channel~1, the heartbeat period for each node is three time slots. A node must transmit an uplink frame in a heartbeat slot. This open a downlink window to maintain the clock synchronization of the node (cf.~\S\ref{subsubsec:clock-sync}) and send active network commands. The heartbeat period can be set according to the nodes' clock drift rates and the required clock accuracy to avoid TDMA panic. The heartbeats also help the LoRaCP controller be aware of whether a node is still alive.

\noindent
{\bf Negative acknowledgment (NAK)}: To deal with frame losses, acknowledging all concurrent uplink transmissions is wasteful because of the downlink-uplink asymmetry. Thus, LoRaCP uses the NAK scheme. In LoRaWAN, the uplink and downlink frames from/to a node have continuously increasing counters, respectively. Thus, both the controller and the nodes can detect if there are lost frames by checking the continuity of the frame counters. If the controller detects lost frames, it sends an NAK using the subsequent downlink transmission to notify the node, which can then use the urgent channel or wait for the next TDMA slot to resend the lost data. The node can also send NAK using the urgent channel or the next TDMA slot to request lost frames. With the NAK scheme, the controller needs not respond to a node's network report if there is no network commands for the node and no lost frames. This design mitigates the contention for the downlink time.

\subsection{Software Architectures of LoRaCP Node and Controller}
\label{subsec:loracp-prototype}

\begin{figure}
  \centering
  \includegraphics{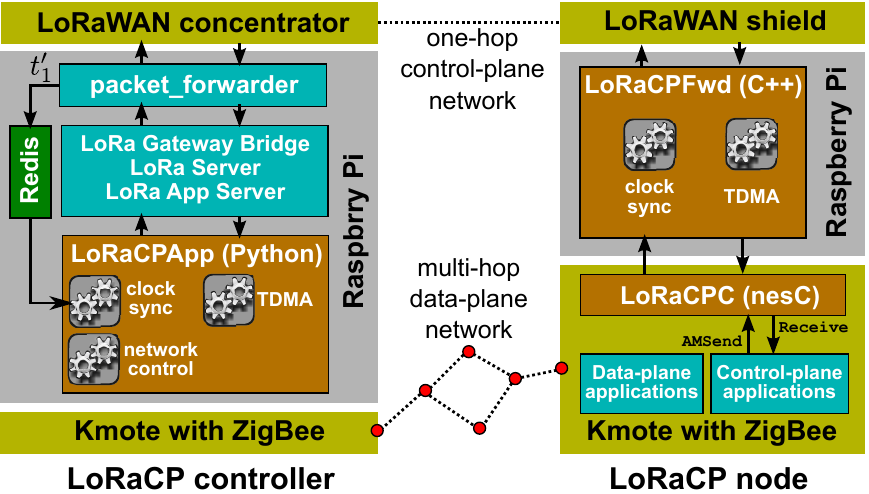}
  \caption{Software architectures of LoRaCP controller and node. (The illustration includes a ZigBee radio for the controller to be a control-plane sink.)}
  \label{fig:software}
\end{figure}

In \S\ref{subsec:lorawan-profiling}, we have introduced the hardware prototypes of the LoRaCP node and controller. This section presents their software architectures as illustrated in Fig.~\ref{fig:software}.

\subsubsection{LoRaCP node}

A C++ forwarder program {\em LoRaCPFwd} runs on the RPi to buffer and forward the data between Kmote and the LoRaWAN shield, while following LoRaCP-MAC. The node parts of the clock synchronization and TDMA are also implemented in {\em LoRaCPFwd}. The Kmote runs TinyOS. We design a TinyOS module {\em LoRaCPC} that provides the {\em AMSend} and {\em Receive} interfaces to send and receive data to/from the RPi through serial communications. Thus, in our prototype design, the Kmote uses LoRaWAN as a service.

\subsubsection{LoRaCP controller}
The RPi of the controller runs an open-source LoRaWAN server architecture \cite{lora-server} consisting of {\em packet\_forwarder}, {\em LoRa Gateway Bridge}, {\em LoRa Server}, and {\em LoRa App Server}. This architecture, through providing JSON-based APIs to subscribe/send messages from/to the LoRaWAN network, greatly simplifies the design of centralized network control applications. The role of this architecture is similar to that of an SDN controller platform (e.g., OpenDaylight) that facilitates the design of SDN control applications. In this paper, the centralized network controls and the controller parts of the clock synchronization and TDMA are implemented in a single Python program called {\em LoRaCPApp}. Note that the LoRaWAN server architecture \cite{lora-server} supports multiple LoRaWAN gateways. Although this paper focuses on a single LoRaCP controller, in the future work, the multi-gateway support can be exploited to develop redundant LoRaCP controllers to improve the system's reliability against a single point of failure.

\subsection{Implementation of LoRaCP Components}

This section provides implementation details of LoRaCP's clock synchronization and TDMA.

\subsubsection{Clock synchronization}
\label{subsubsec:clock-sync}

Clock synchronization is a basis for implementing TDMA. Although there are various existing clock synchronization protocols for LMWNs (e.g., FTSP), if we synchronize the LoRaCP nodes to the controller using the data-plane network, the control plane's TDMA will depend on the data-plane network, incurring the undesirable coupling. Thus, we synchronize the LoRaCP nodes to the controller using the control-plane network. However, there is still limited research on clock synchronization over LoRaWAN. In our prototype system, the RPi's clock is used as the node's or controller's clock. Although the LoRaWAN devices and the Kmote have their own timers, using the RPi's clock can simplify the evaluation of the accuracy of the LoRaWAN clock synchronization using the RPi's Ethernet interface.

To save the downlink time, LoRaCP does not prescribe dedicated frames for clock synchronization. Instead, LoRaCP piggybacks several bytes to each control-plane frame for clock synchronization. Specifically, each uplink frame is appended with the node's clock value $t_0$ as illustrated in Fig.~\ref{fig:lora_ntp}. The controller records its clock value $t_1'$ on completion of the frame reception. The clock offset between the node and the controller, denoted by $\delta$, can be estimated as $\delta = t_1' - (t_0 + \Delta)$, where $\Delta$ is the uplink latency presented in Fig.~\ref{fig:on_air_time}. Then, the controller piggybacks $\delta$ onto the downlink frame as illustrated in Fig.~\ref{fig:lora_ntp}. Upon receiving $\delta$, the node resets its clock by $t = t + \delta$, where $t$ denotes the node's current clock value. Alternatively, the node's clock advance speed can be calibrated according to $\delta$ using a negative feedback loop.

We now discuss several implementation issues of the above clock synchronization approach. First, the LoRaWAN frame header added by the shield has changeable size because the integers in the headers are represented as variable length hexadecimal ADSII strings. As shown in Fig.~\ref{fig:on_air_time}, the uplink latency $\Delta$ has a complex relationship with the frame size in different channels. When the LoRaCP controller receives the uplink frame, it checks the actual frame size and the SF used by the node to query the corresponding $\Delta$ from the data in Fig.~\ref{fig:on_air_time}. Thus, for LoRaWAN clock synchronization, the prior knowledge in Fig.~\ref{fig:on_air_time} is critical. Note that most LMWN clock synchronization approaches are free from this frame size dependence issue because they use dedicated synchronization frames with fixed sizes or the frame size has little impact on transmission latency. Second, we modify {\em packet\_forwarder}, i.e., LoRaWAN concentrators' driver program, to record $t_1'$, because other components of the LoRaWAN server architecture may suffer software delays. As illustrated in Fig.~\ref{fig:software}, the timestamp $t_1'$, together with the corresponding source ID and frame ID, are written into a Redis in-memory database and then retrieved by the {\em LoRaCPApp} to compute $\delta$.

We measure the synchronization accuracy of the above approach using the \texttt{ntpdate} tool to check the clock offset between the node and the controller over a local Ethernet network connecting the RPis. The mean absolute synchronization error is $2.9\,\text{ms}$ with a standard deviation of $1.7\,\text{ms}$. Given the second-level frame transmission time, such synchronization errors of a few milliseconds are satisfactory.

\subsubsection{TDMA}

The prototype LoRaCP node controls the sleep of the LoRaWAN radio and transmissions of frames based on its RPi's synchronized clock. Specifically, if {\em LoRaCPFwd} has received a network report from the Kmote, the RPi starts awaking the LoRaWAN radio $850\,\text{ms}$ before its next TDMA time slot, transmits the report in the time slot, receives any subsequent network command, re-transmits frames using the urgent channel if an NAK is received. Finally, {\em LoRaCPFwd} forwards all received network commands to the Kmote. In our current experimental implementation, we assign LoRaWAN channels and time slots to nodes manually. The non-essential operations such as the automatic channel and time slot assignments, support of adding and dropping nodes at run time, etc, are left to future work.
\section{Performance Evaluation}
\label{sec:eval}

\subsection{Experiment Methodology and Settings}

We use LoRaCP to implement the CTP-SCDP discussed in \S\ref{subsec:distributed-vs-centralized}. Specifically, if the Kmote of a LoRaCP node detects a change of ETX with any of its neighbor node, it uses the {\em LoRaCPC} to send the latest ETX using an network report frame to the LoRaCP controller. Upon receiving the ETX update, the controller's {\em LoRaCPApp} python program computes the optimal routing and pushes network commands containing new parent node information to the downlink queue of the LoRaWAN server architecture. Upon receiving a network command, a LoRaCP node updates its parent node accordingly. In the data plane, each node generates a data packet every eight seconds.

We conduct experiments on a testbed consisting of a LoRaCP controller and 15 LoRaCP nodes. The nodes are placed at the grid points of a lab space. The nodes are evenly divided to use three LoRaWAN channels (SF7, SF8, and SF9). The time slot lengths in these three channels are 3, 4, and 5 seconds, respectively. The controller uses the first downlink window RX1 to transmit network commands. Before the RX1 window, the controller has a {\em wait time} of one second to compute the network commands, which is generally sufficient. On our 16-node testbed, each LoRaCP has a time slot every 25 seconds or less. For larger networks, to maintain this rotating period for each node, multiple geographically distributed nodes in the same channel can be assigned to use the same time slot, since they unlikely report ETX changes at the same time. We leave the evaluation of this extended LoRaCP-MAC to future work after we expand our testbed.

\subsection{Experiment Results}

We conduct three sets of experiments: \S\ref{subsubsec:pressure} pressure-tests LoRaCP; \S\ref{subsubsec:ctp-scdp-control} evaluates the control plane performance of CTP-SCDP; \S\ref{subsubsec:ctp-scdp-data} compares CTP and CTP-SCDP.

\subsubsection{Control plane pressure tests}
\label{subsubsec:pressure}

\begin{figure}
  \subfigure[Downlink delay.]
  {
    \includegraphics{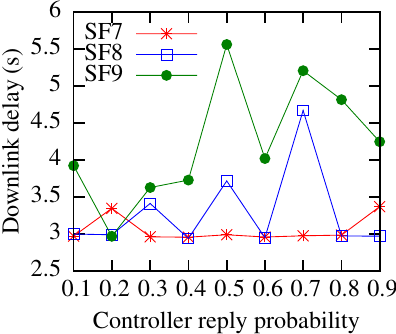}
    \label{fig:tdma-delay}
  }
  \subfigure[Downlink FDR.]
  {
    \includegraphics{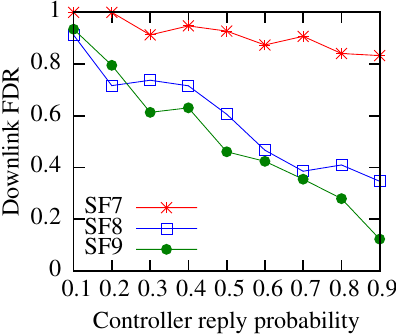}
    \label{fig:tdma-pdr}
  }
  \caption{Control plane pressure test results.}
  \label{fig:downlink-perf}
\end{figure}

While the concurrent uplink channels increase the throughput for network reports, LoRaWAN's downlink-uplink asymmetry presents a bottleneck for the downlink communications. We pressure-test the downlink performance. Specifically, each LoRaCP node transmits a network report every its time slot. Thus, the controller receives frames from the three channels concurrently almost at all the time. It replies to each network report with a certain probability. The frame size of the replies ranges from 29 to 33 bytes. NAK is turned off in these tests.

Fig.~\ref{fig:downlink-perf} shows the average control-plane downlink delays and frame delivery ratios (FDRs) of different channels versus the probability that the controller replies. The downlink delay is measured as the time duration between i) the controller's {\em LoRaCPApp} pushes a network command to the LoRaWAN server architecture and ii) the node's {\em LoRaCPFwd} receives the command. This downlink delay includes the wait time of one second. From Fig.~\ref{fig:tdma-delay}, the average downlink delay does not significantly increase with the controller's reply probability. The average delay ranges from $3\,\text{s}$ to $5.5\,\text{s}$. It increases with the SF, because a larger SF has a lower data rate. Fig.~\ref{fig:tdma-pdr} shows the control-plane downlink FDR versus the controller's reply probability. The FDR decreases with the reply probability. This is because the open-source LoRaWAN server architecture \cite{lora-server} drops frames when it receives excessive frames to be transmitted beyond the downlink throughput. From the results in Fig.~\ref{fig:downlink-perf}, the downlink bottleneck mainly affects the downlink FDR. Thus, in the remaining experiments, we use the downlink FDR to assess whether the control plane performance is throttled by the downlink-uplink asymmetry.

\subsubsection{Control plane performance in CTP-SCDP}
\label{subsubsec:ctp-scdp-control}

\begin{figure}
  \subfigure[Per-node energy consumption for the control plane in one hour.]
  {
    \includegraphics{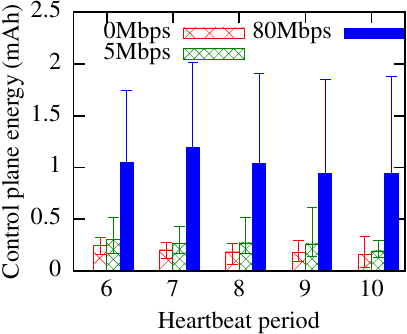}
    \label{fig:fig:interference-energy}
  }
  \subfigure[Downlink FDR.]
  {
    \includegraphics{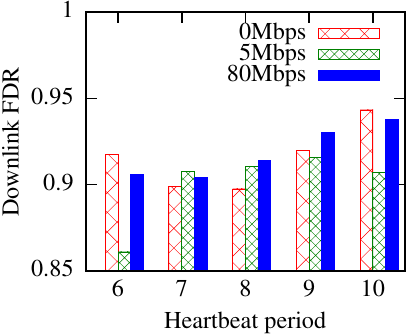}
    \label{fig:interference-pdr}
  }
  \caption{CTP-SCDP control plane performance under Wi-Fi interference against the data-plane network. The error bar represents min and max values.}
  \label{fig:interference}
\end{figure}

We evaluate the performance of CTP-SCDP's control-plane network. To create data-plane link quality variations, we use a laptop placed close to the testbed to generate Wi-Fi traffic to interfere with the ZigBee data-plane network. ZigBee radios use Channel 18 and the Wi-Fi AP uses Channel 6, which interfere with each other. On the laptop, we use \texttt{iperf3} to generate data traffic at a specified bit rate. This experiment methodology well captures the increasingly crowded $2.4\,\text{GHz}$ ISM band used by the ZigBee-/BLE-based data-plane networks. In the presence of the Wi-Fi interference, the CTP-SCDP generates more control-plane messages to report the volatile link ETXes of the data-plane network to the LoRaCP controller.

First, we estimate the energy consumption of each LoRaCP node's LoRaWAN shield by multiplying the transmitting/receiving currents with the measured total times in respective modes. Fig.~\ref{fig:fig:interference-energy} shows the error bars of per-node energy consumption by the shield in one hour under different settings of heartbeat period and Wi-Fi interference intensity. The control-plane energy consumption increases with the interference intensity due to the increased control-plane messages. When we do not generate Wi-Fi interference, the energy consumption decreases with the heartbeat period. This is because, in the absence of the interference, the link ETXes seldom change and most control-plane messages are the heartbeats. In the presence of interference (i.e., $5\,\text{Mbps}$ and $80\,\text{Mbps}$), the energy consumption has no monotonic relationship with the heartbeat period, because the node will utilize the non-heartbeat time slots to report the volatile ETXes. From Fig.~\ref{fig:fig:interference-energy}, with no and intensive interference ($80\,\text{Mbps}$), the per-node power consumption by the control plane averaged over time is about $0.825\,\text{mW}$ and $3.3\,\text{mW}$, respectively, which are comparable to or lower than the power consumption of low-power microcontrollers (MCUs). For instance, the active power of TelosB's MCU is $5.94\,\text{mW}$, whereas the recent Firestorm's MCU consumes $28.38\,\text{mW}$ in the common configuration \cite{andersen2016system}.

Second, we measure the average control-plane downlink FDR over all channels. The results are shown in Fig.~\ref{fig:interference-pdr}. Even if the data-plane network experiences intensive interference, the FDR is generally above 90\%. Thus, the CTP-SCDP's control plane is still beyond the downlink bottleneck.

\subsubsection{Comparison between CTP and CTP-SCDP}
\label{subsubsec:ctp-scdp-data}

\begin{figure}
  \subfigure[Data-plane PDR and control-plane downlink FDR.]
  {
    \includegraphics{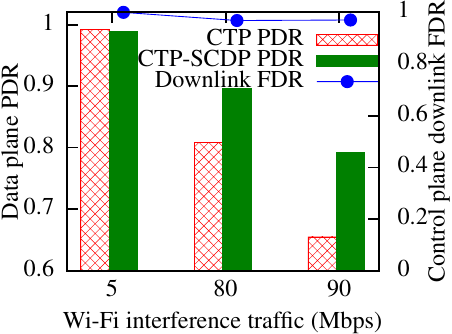}
    \label{fig:side-by-side-pdr}
  }
  \subfigure[Control-plane uplink frames and per-node energy in one hour.]
  {
    \includegraphics{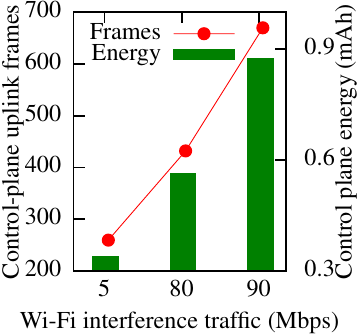}
    \label{fig:control-plane-energy}
  }
  \vspace{-1em}
  \caption{Performance comparison between CTP and CTP-SCDP.}
  \label{fig:ctp-vs-ctp-scdp}
\end{figure}

We load CTP to eight nodes and CTP-SCDP to another eight nodes. We run CTP and CTP-SCDP side by side on the testbed, so that they experience almost the same Wi-Fi interference for fair comparison. CTP-SCDP's LoRaCP heartbeat period is 10. Fig.~\ref{fig:side-by-side-pdr} shows the data plane's packet delivery ratio (PDR), i.e., the ratio of the ZigBee packets received by the data-plane sink over all packets generated by the source nodes. When the Wi-Fi interference intensity is low ($5\,\text{Mbps}$), CTP and CTP-SCDP achieve similarly high PDRs. When the interference intensity is $80\,\text{Mbps}$, CTP-SCDP's PDR is 10\% higher than CTP's. When the interference intensity is $90\,\text{Mbps}$, CTP's PDR drops to 65\%, while CTP-SCDP's is 80\%. Note that the actual data rate of the Wi-Fi interference traffic fluctuates over time. Moreover, the fluctuation level increases with the setpoint. The data rate deviations are $0.8\,\text{Mbps}$ only and up to $20\,\text{Mbps}$ for setpoints $5\,\text{Mbps}$ and $90\,\text{Mbps}$, respectively. Thus, the control-plane networks experience more dynamic interference with a larger setpoint, resulting in the increasing PDR gain of CTP-SCDP over CTP with the interference intensity setting. This result is consistent with our observation from the simulation study in \S\ref{subsec:distributed-vs-centralized} that CTP cannot handle dynamic network conditions well.

Fig.~\ref{fig:side-by-side-pdr} also shows the control-plane downlink FDRs, which are above 97\%. This suggests that the control plane is beyond the downlink bottleneck. Fig.~\ref{fig:control-plane-energy} shows the total number of control-plane uplink frames of CTP-SCDP during one hour and the projected per-node energy consumption by the LoRaWAN shield. In the presence of stronger interference, more uplink frames will be transmitted to report the volatile ETXes. With $5\,\text{Mbps}$ and $90\,\text{Mbps}$ interference, the total numbers of data-plane transmissions (including beacons and forwarded packets) are 5,022 and 10,024, respectively. The corresponding numbers of control-plane uplink frames are just 5.2\% and 6.7\% of these data-plane transmissions. With strong interference ($90\,\text{Mbps}$), the per-node control-plane power consumption averaged over time is less than $2.97\,\text{mW}$, consistent with the results in Fig.~\ref{fig:interference} obtained with 15 nodes.
\section{Conclusion and Future Work}
\label{sec:conclude}

This paper studied using LoRaWAN radios to form one-hop out-of-band control planes for LMWNs through extensive measurement study, system design, and testbed evaluation. We demonstrated applying the designed system, LoRaCP, to physically separate the control plane of CTP from its ZigBee-based data-plane network. Experiments show that LoRaCP increases CTP's packet delivery ratio from 65\% to 80\% in the presence of external interference, while additionally consuming a per-node average radio power of $2.97\,\text{mW}$. In future work, we will evaluate systematically the network performance and manageability gains by LoRaCP, as well as its impact on node lifetime in real-world environments such as factories.



\bibliographystyle{./IEEEtran}
\bibliography{./IEEEabrv,./bibtex}

\end{document}